\documentclass[10pt,a4paper,twocolumn, twoside]{article}
\RequirePackage{graphicx}
\RequirePackage[switch]{lineno}
\RequirePackage{color}
\RequirePackage[colorlinks,citecolor=blue,urlcolor=blue,linkcolor=blue]{hyperref}
\usepackage{graphics}
\usepackage{graphicx} 
\usepackage{dcolumn}
\usepackage{bm}
\usepackage{multirow}
\usepackage{upgreek}
\usepackage{longtable}
\usepackage{booktabs}
\usepackage[latin1]{inputenc}
\usepackage{caption}
\usepackage[affil-it]{authblk}
\usepackage{blindtext}
\usepackage{abstract}
\usepackage{amsmath,amssymb}

\makeatletter
\renewcommand{\section}{\@startsection{section}{2}{0cm}{-\baselineskip}
{0,5\baselineskip}{\normalsize\bfseries}}
\renewcommand{\subsection}{\@startsection{subsection}{3}{0cm}{-\baselineskip}
{0,5\baselineskip}{\normalsize\slshape}}
\makeatother

\begin{document}

\title{Constraints on elastic neutrino nucleus scattering in the fully coherent regime from the \textsc{Conus} experiment}

\author{H.~Bonet$\rm ^1$, A.~Bonhomme$\rm ^1$, C.~Buck$\rm ^1$, K.~F\"ulber$\rm ^2$, J.~Hakenm\"uller$\rm ^1$, G.~Heusser$\rm ^1$,  T.~Hugle$\rm ^1$, M.~Lindner$ \rm ^1$,  W.~Maneschg$\rm ^1$, T.~Rink$\rm ^1$, H.~Strecker$\rm ^1$, R.~Wink$\rm ^2$\\
\textsc{Conus} {\rm Collaboration}}


\date{\small \it 
$^1$Max-Planck-Institut f\"ur Kernphysik, Saupfercheckweg 1, 69117 Heidelberg, Germany \\
$^2$Preussen Elektra GmbH, Kernkraftwerk Brokdorf, Osterende, 25576 Brokdorf, Germany\\
\vspace{0.3cm}
E-mail address: \\ {\tt conus.eb@mpi-hd.mpg.de} 
\vspace{0.3cm}
} 

\twocolumn[
\begin{@twocolumnfalse}
\maketitle

\begin{abstract}
We report the best limit on coherent elastic scattering of electron antineutrinos emitted from a nuclear reactor off germanium nuclei. The measurement was performed with the \textsc{Conus} detectors positioned at 17.1\,m from the 3.9\,GW$_{th}$ reactor core of the nuclear power plant in Brokdorf, Germany. The antineutrino energies of less than 10\,MeV assure interactions in the fully coherent regime. The analyzed dataset includes 248.7\,kg$\cdot$d with the reactor turned on and background data of 58.8\,kg$\cdot$d with the reactor off. With a quenching parameter of $k=0.18$ for germanium, we determined an upper limit on the number of neutrino events of 85 in the region of interest at 90\% confidence level. This new CONUS dataset disfavors quenching parameters above $k=0.27$, under the assumption of standard-model-like coherent scattering of the reactor antineutrinos.
\\
\noindent {\it Keywords:} Nucleus-neutrino interactions, Semiconductors, Nuclear reactors\\
\end{abstract}
\end{@twocolumnfalse}
]
\vspace{1.0cm}

Coherent elastic scattering of neutrinos off nuclei (CE$\nu$NS) has been predicted on the basis of the standard model (SM) since 1974~\cite{Freedman:1973yd}. Compared to other neutrino interaction channels, CE$\nu$NS is appealing because of its larger cross section. It scales with the squared number of neutrons in the target nucleus and the squared neutrino energy. In principle, this enhancement allows for the design and construction of small-sized neutrino detectors. However, the coherence condition of the momentum transfer being smaller than the inverse of the nuclear size, does not hold for neutrino energies above a few tens of MeV. Moreover, the observables are low energy nuclear recoils, which are suppressed with increasing atomic number. Thus, coherent scattering eluded detection for many decades. Only with the recently emerging low energy threshold technologies~\cite{Barbeau:2007qi, PhysRevD.91.072001} has it become approachable. The process of CE$\nu$NS was first observed by the COHERENT Collaboration in 2017 using a CsI[Na] scintillator~\cite{Coherent:2017} followed by a detection on argon~\cite{Coherent:2020}. For both measurements, the COHERENT experiment utilized neutrinos produced via pion decays at rest in the spallation neutron source of the Oak Ridge National Laboratory. Instead, the \textsc{Conus} experiment aims for a complementary observation of CE$\nu$NS in the fully coherent regime with lower energetic neutrinos~\cite{Kerman:2016jqp, Bednyakov:2018mjd} emitted by a nuclear reactor. Worldwide, similar projects at nuclear reactors are planned or ongoing~\cite{Singh:2017jow, Aguilar-Arevalo:2019zme, Strauss:2017cuu, Agnolet:2016zir, Belov_2015, Akimov:2019ogx}.

The study of CE$\nu$NS events provides a probe to SM and beyond the standard model physics. For instance, it gives direct access to the neutron form factor and thus nucleon density~\cite{Amanik:2009zz}. Beyond that, this channel allows one to probe deviations of the expected weak mixing angle at the MeV scale~\cite{Kosmas:2015vsa}, the existence of nonstandard neutrino-quark interactions~\cite{Barranco:2005yy}, and light new states~\cite{deNiverville:2015mwa}, as well as electromagnetic properties such as a neutrino magnetic moment~\cite{Kosmas:2015vsa, Beda:2013mta} or an effective charge radius~\cite{Papavassiliou:2005cs}. The high antineutrino flux at reactor sites might allow one to investigate the reactor antineutrino rate anomaly~\cite{Mention:2011} via the CE$\nu$NS channel. New insights on sterile neutrinos~\cite{Gariazzo:2015rra, Boser:2019rta} or the origin of the anomalous features observed in reactor antineutrino spectra~\cite{Abe:2014} could be gained. The development of compact neutrino-sensitive devices will further enable the monitoring of the reactor thermal power and the proliferation of fissile products such as plutonium~\cite{Bernstein:2019hix}. Finally, the understanding gained with these new detector technologies can be relevant for astrophysics and cosmology. For example, CE$\nu$NS is expected to play a major role in supernovae, i.e.~in the cooling and heating processes during a stellar collapse~\cite{Drukier:1984}. In addition, solar, diffuse supernovae remnant and atmospheric neutrinos interact via CE$\nu$NS in dark matter detectors producing an irreducible background (`neutrino floor'). A precise measurement of the CE$\nu$NS cross section will allow for a better prediction of this background.
 
Nuclear reactors are known to be strong sources of electron antineutrinos with energies below 10\,MeV. As continuous and well-localized sources, they offer an ideal environment to investigate neutrino properties. The \textsc{Conus} experiment is carried out at the commercial nuclear power plant in Brokdorf (KBR), Germany, which is operated by the Preussen Elektra GmbH. The design of this pressurized water reactor corresponds to a single unit power station featuring a reactor core of 193 fuel assemblies, with a length of 3900\,mm of the active zone. The maximum thermal power amounts to 3.9\,GW$_{th}$, which generates a gross electric power of 1.48\,GW. The time-dependent thermal power is calculated via three different methods: the thermal balance in the secondary circuit, ionization chambers around the reactor core, and a core simulation. This thermal power information is available in intervals of 2 hours or less, with an uncertainty of 2.3\% (1$\sigma$)~\cite{Hakenmuller:2019ecb}. Besides the thermal power, the average energy released per fission needs to be known to predict the antineutrino flux emitted by the reactor. This information is taken for the most relevant fissile nuclides $^{235}$U, $^{239}$Pu, $^{238}$U, and $^{241}$Pu from~\cite{Ma:2012bm}. The run-specific antineutrino spectra were obtained by summation of the predictions from Huber ($^{235}$U, $^{239}$Pu and $^{241}$Pu)~\cite{Huber:2011wv} and Mueller ($^{238}$U)~\cite{Mueller:2011nm} taking into account the average fission fractions of the fissile nuclides at KBR. Correction factors were applied as determined by the Daya Bay experiment~\cite{DBflux}. The uncertainty on the reference spectra is based on the covariance matrix provided by the Daya Bay Collaboration~\cite{DBflux}.

The \textsc{Conus} experiment is located inside the safety containment of KBR (room A-408). The basic infrastructure comprises electricity and pressurized air connectors, as well as cold air ventilation that keeps the room temperature constant~\cite{conus-det}. The experimental setup is positioned directly beneath a pool, which contains spent fuel assemblies immersed in cooling water with a level of ($13.4\pm0.1$)\,m. Contributions to the expected signal coming from neutrons~\cite{Hakenmuller:2019ecb} or neutrinos produced in these spent fuel assemblies can be neglected. Together with the thick ferroconcrete layers inside the containment sphere, the overburden amounts to $10-45$\,m of water equivalent (m~w.e.), depending on the azimuth angle. On average it is 24\,m~w.e., which is large enough to shield efficiently against the hadronic component of cosmic rays at surface. The location of the \textsc{Conus} detectors is only 17.1\,m from the reactor core center, leading to an antineutrino flux of 2.3$\cdot$10$^{13}$\,cm$^{-2}\cdot$s$^{-1}$ at maximum thermal power. Room A-408 is outside the innermost biological shield, which surrounds the reactor core, and thus accessible at any time. A more detailed description of the experimental site and the parts of the reactor geometry relevant for \textsc{Conus} is given in~\cite{Hakenmuller:2019ecb}.

The \textsc{Conus} setup consists of four p-type point contact high purity germanium (HPGe) detectors (in the following denoted C1 to C4)~\cite{conus-det}. Each cylindric germanium (Ge) diode has a mass of 0.996\,kg. Considering dead layers, this leads to a total fiducial mass of ($3.73\pm0.02$)\,kg. The novel detector design had to fulfill five key requirements, which were elaborated and met in close cooperation with Mirion Technologies (Canberra) in Lingolsheim, France. First, ultralow noise levels had to be achieved in order to obtain pulser resolutions better than 85\,eV$_{ee}$ (electron equivalent energy) FWHM. Second, cosmic activation of the freshly pulled Ge crystals and other detector parts had to be minimized. Third, all components had to undergo a strict radiopurity selection. Fourth, electrically powered cryocoolers (Canberra Cryo-Pulse 5 Plus) had to be used, since for safety reasons at reactor site liquid nitrogen dewars are not allowed to refrigerate the Ge diodes. Fifth, the cryostat arm lengths had to be exceptionally long for these detectors in order to be compatible with the dimensions of the shield design. 

\begin{figure}
	\includegraphics[width=0.5\textwidth]{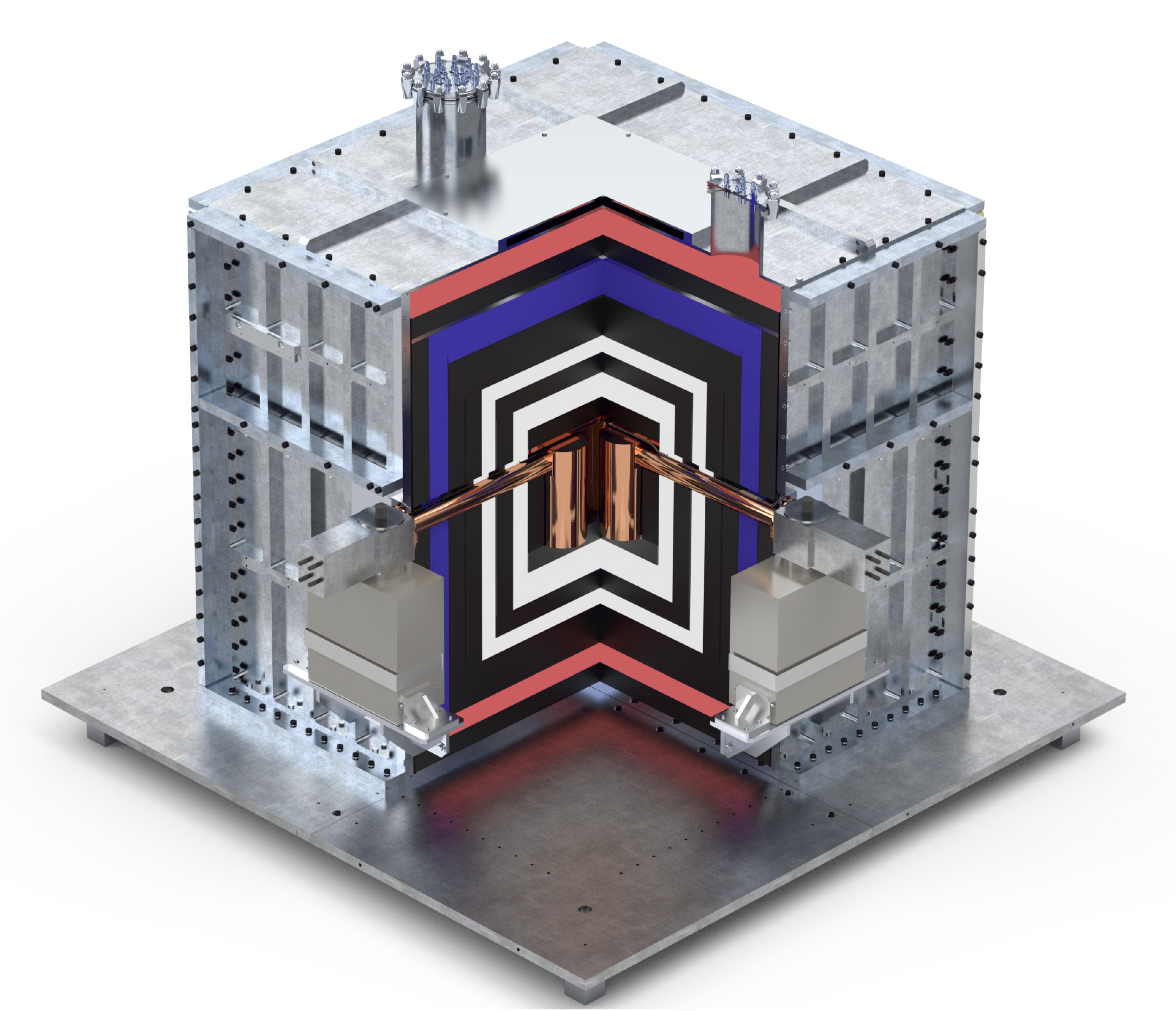}
	\caption{\textsc{Conus} setup. Shield layers consist of steel (silver), PE (red), Pb (black), B doped PE (white) and plastic scintillator (blue) used as muon anticoincidence system. In the center, four HPGe detectors are embedded in ultralow background Cu cryostats and connected to electrically powered cryocoolers.}
	\label{fig:1}
\end{figure}

\textsc{Conus} employs an onionlike shield, which is depicted in Figure~\ref{fig:1}. It is based on decades-long research and development programs for highly sensitive Ge $\gamma$-ray spectrometry at MPIK (e.g.~\cite{Heusser:1995wd}), in particular on the GIOVE detector~\cite{Heusser:2015ifa} at shallow depth (15\,m~w.e.). Five layers of lead (Pb), in total 25\,cm with increasing radiopurity toward the center, weaken the external $\gamma$-radiation by at least five orders of magnitude. The innermost layer consists of radiopure Pb with a mean specific $^{210}$Pb activity of $<1.1$\,Bq/kg. Compared to copper (Cu), Pb is advantageous for CE$\nu$NS detection, since the residual muon-induced bremsstrahlung continuum at low energies has a lower intensity due to a stronger self-absorption. One layer of plastic scintillation plates of type EJ-200~\cite{EJ200} was installed as an active muon anticoincidence system. Herein, each side module was equipped with two photomultiplier tubes (PMTs) and the top module with four PMTs (Hamamatsu R11265 U-200), one on each corner. With a gate length of 410\,$\upmu$s it allows one to reject $\sim$97\% of prompt muon-induced signals. Two layers of boron (B) doped polyethylene (PE) are shifted closer to the innermost shield layers than for GIOVE. This was possible, since the B compound in \textsc{Conus}, boric acid highly enriched in $^{10}$B, was comparatively more radiopure, in particular with respect to $^{40}$K. These layers are meant to efficiently shield neutrons, either created by muons in Pb or originating from the reactor core. The setup is encapsulated by a steel frame, which guarantees safety requirements to be met and helps reduce radon (Rn) diffusion into the detector chamber. With a volume of 1.65\,m$^3$ and a total mass of 11\,tons, the \textsc{Conus} shield is extremely compact. The background suppression capability of the \textsc{Conus} shield was tested at MPIK next to the GIOVE setup. First, the CONus RAdiation Detector (CONRAD) was installed. It is an ultralow background Ge detector from the Genius-TF experiment~\cite{Baudis:2002} with a fiducial mass of ($1.9\pm0.2$)\,kg. The results were compared with GIOVE data and used to validate Monte Carlo (MC) simulations of natural and cosmogenic background. Then, the \textsc{Conus} detectors were stepwise integrated and the new background conditions monitored accordingly.

At MPIK as well as at KBR, the dominant overall background source in the \textsc{Conus} detectors was identified to be events induced by the interactions of cosmic ray muons inside the shield or the surrounding materials. This kind of background does not correlate with the reactor power and can be measured during reactor OFF periods. Seasonal or barometric effects in the muon flux do not create significant differences in the background level of the ON/OFF datasets. Another reactor-uncorrelated background source are signals from Rn progenies. The Rn concentration in air has been continuously monitored and determined to be (175$\pm$35)\,Bq/m$^3$. The \textsc{Conus} shield is flushed using breathing air bottles, that have been stored for at least three weeks. In this period, most of the originally present Rn has decayed. A more critical potential source of background are reactor-correlated neutrons, whose recoils could mimic CE$\nu$NS signals. Precise neutron spectrometry measurements with the well-calibrated NEMUS setup~\cite{WIEGEL200236} were performed by department 6.4 (neutron radiation) of the Physikalisch-Technische Bundesanstalt (PTB) Braunschweig, Germany. Together with MC simulations of the reactor neutron propagation and auxiliary $\gamma$-ray measurements it was possible to demonstrate that the thermal-power-correlated neutron field adds only a negligible contribution to the background budget of the \textsc{Conus} detectors~\cite{Hakenmuller:2019ecb}.

The background is simulated by MC simulations. Excellent agreement between data and MC simulations is observed at high energies around 100\,keV$_{ee}$ and above. Toward lower energies the most relevant contributions to the background are events induced by cosmic ray muons, activation lines from Ge, $^{210}$Pb on surfaces inside the cryostat and Rn~\cite{Hakenmuller:phd}. The background level slightly above the expected CE$\nu$NS signal is on the order of 10\,counts kg$^{-1}\cdot$d$^{-1}\cdot$keV$^{-1}$. Data and MC simulations confirmed background variations with time to have negligible impact on the analysis. Below 0.5\,keV$_{ee}$ toward the energy threshold of the analysis, an additional component starts to contribute, which is associated with electronic noise, not included in the MC model of the physical background. 

\begin{figure}
	\includegraphics[width=0.5\textwidth]{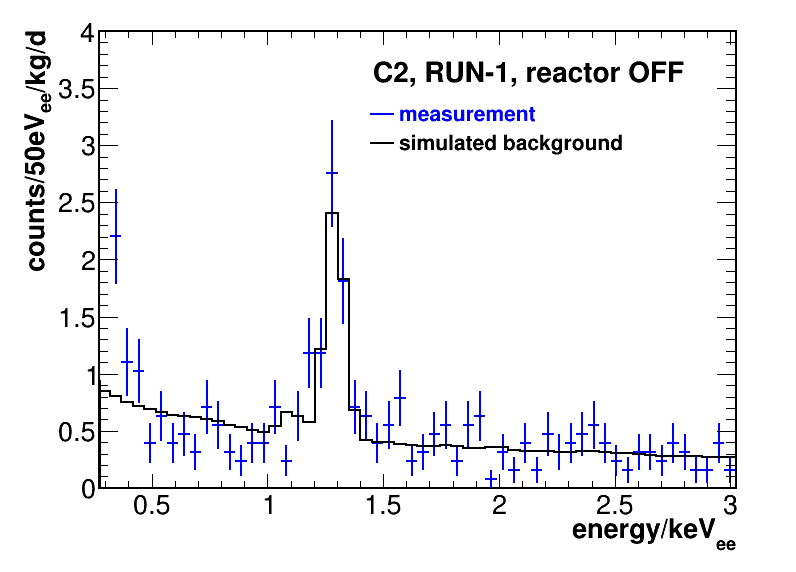}
	\caption{Comparison of background model to experimental data for the C2 detector. At low energies an increase of the measured count rate as compared to the model due to the electronic noise component is observed.}
	\label{fig:2}
\end{figure}

Within statistics, good agreement between the background model and the reactor OFF data is obtained for energies above the electronic noise contribution as shown exemplarily in Figure~\ref{fig:2}. This background model from MC simulations is used as input to the likelihood analysis described below. 

\textsc{Conus} started data collection during a reactor outage in April 2018. The dataset analysed in this publication includes data with the reactor ON between May 2018 and June 2019. An additional reactor OFF period in June-July 2019 allowed us to measure background. The data are divided in several run periods defined by different data acquisition (DAQ) settings and noise levels~\cite{conus-det}. This analysis includes data of the first two \textsc{Conus} runs. Each of the runs contains its own reactor OFF period as shown in Table~\ref{tab:1}. Stability of environmental data was a basic criterium for the definition of the datasets. In particular, reactor ON intervals were only included in the analysis, if the temperature and the noise level were comparable to the associated reactor OFF interval~\cite{conus-det}. For C4 a temporarily appearing artifact was observed in the spectrum during RUN-1, therefore this dataset was excluded. Moreover, based on the time difference distribution of events~\cite{conus-det}, noise induced by the cryocoolers with characteristic frequencies and microphonics was removed without loss of efficiency for the physical signals. The total reactor ON run time corresponds to 248.7\,kg$\cdot$d and the total reactor OFF run time to 58.8\,kg$\cdot$d.

\begin{table}
\caption{\label{tab:1}Live times for reactor ON and OFF periods as well as the regions of interest for the \textsc{Conus} detectors used in the analysis.}
\centering
\begin{tabular}{lcccc}
Det. & RUN & ON [d] & OFF [d] & ROI [keV$_{ee}$] \\\hline
C1 & 1 & 96.7 & 13.8 & 0.296 - 0.75 \\
C2 & 1 & 14.6 & 13.4 & 0.311 - 1.00 \\
C3 & 1 & 97.5 & 10.4 & 0.333 - 1.00 \\\hline
C1 & 2 & 19.6 & 12.1 & 0.348 - 0.75 \\
C3 & 2 & 20.2 & 9.1 & 0.343 - 1.00 \\\hline
\end{tabular}
\end{table} 

The regions of interest were chosen individually for each detector and run period. For the definition of each region of interest (ROI), three criteria were adopted. First the trigger efficiency of the detector must not drop significantly from unity ($>95$\,\%)~\cite{conus-det}. Next, the threshold had to be above energies for which rate correlations to variations in the ambient temperature were observed. The simulated background events are stable independent of the room temperature. Finally, the ratio of the electronic noise contribution to the MC background had to be smaller than a factor four. The last condition was applied to reduce the impact of the noise level close to the energy threshold. Table~\ref{tab:1} summarizes the ROI for the different detectors and run periods. The energy cut at 1\,keV$_{ee}$ was chosen due to x-ray lines around [1.0,1.4]\,keV$_{ee}$. In the case of the C1 detector, a slightly lower value of 0.75\,keV$_{ee}$ was picked due to some mismatch between data and simulation in the region from 0.75 to 1\,keV$_{ee}$, which is well above any predicted end of the CE$\nu$NS signal spectrum. 

\begin{figure}
\begin{center}
	\includegraphics[width=0.4\textwidth]{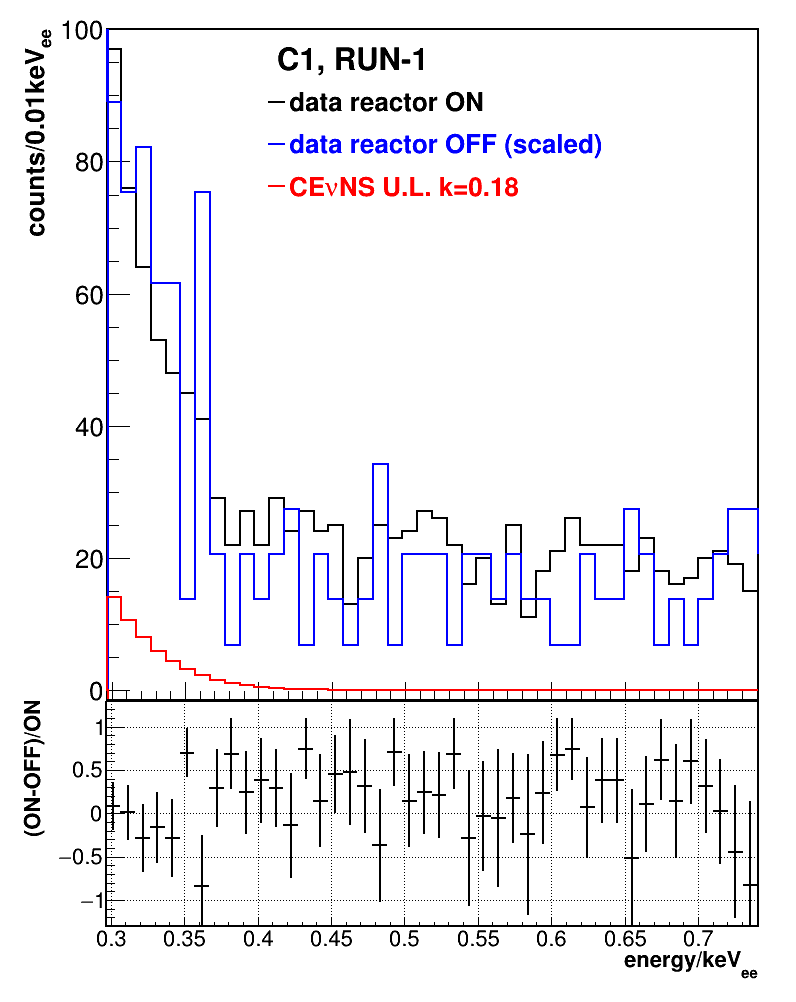}
	\caption{Measured spectra during reactor ON and OFF periods for one of the \textsc{Conus} detectors including weighted differences (bottom). The predicted pure antineutrino spectrum is shown in red for a quenching parameter of 0.18 in case that the signal would be at our 90\% C.L.~limit.}
	\label{fig:3}
\end{center}
\end{figure}

A binned likelihood analysis was employed to extract the amplitude of the CE$\nu$NS signal, via the overall normalization parameter of interest ($s$). The reactor ON and reactor OFF data were fitted simultaneously. The background has three free parameters and consists of the MC simulated model completed by the analytical effective description of the electronic noise edge by an exponential. One of the three parameters provides the overall background normalization ($b$), the other two describe the exponential noise component. Four Gaussian pull terms were added to the likelihood function, allowing one to encode the systematic uncertainties on the energy scale, the normalization (fiducial mass and efficiency) and the neutrino flux. The content of each bin was assumed to follow a Poisson distribution, while an energy binning of 10\,eV$_{ee}$ was chosen. The performed combined fit for the full dataset includes the detectors and runs as listed in Table~\ref{tab:1}. As an example measured spectra for a specific dataset are shown in Figure~\ref{fig:3}. 

The dead time induced by the muon veto was estimated independently for reactor ON and OFF times using different methods based on $^{228}$Th calibration, pulser and veto trigger rate data. It was determined to be on average ($5.8\pm0.2$)\,\% in reactor ON and ($3.5\pm0.2$)\,\% in reactor OFF periods. The DAQ-induced dead time is typically in the percent level~\cite{conus-det}. The stability of the energy scale was confirmed in regular calibration campaigns including pulser scans and deployments of a radioactive $^{228}$Th source. Moreover internal Ge activation lines are used for the energy scale determination~\cite{conus-det}. The energy scale uncertainty of $\sim15$\,eV$_{ee}$ in the ROI is taken into account in the likelihood via one of the pull terms. 

The not yet well-known dissipation processes of nuclear recoils in Ge (quenching) at the cryogenic detector temperatures between 78 and 88\,K are described by the modified Lindhard theory~\cite{lindhard1963range, Barker:2012ek} including an adiabatic correction and the free parameter $k$. This parameter $k$ corresponds roughly to the quenching factor at a nuclear recoil energy of 1\,keV. By performing the fit for different values of $k$ from 0.1 to 0.3, we account for the full spread of measured quenching values at low energy found in literature~\cite{Barker:2012ek, Scholz:2016} as well as a potential diode temperature dependence. The predicted event rate has a strong $k$ parameter dependence, and it changes by more than one order of magnitude within the relevant region. 

\begin{figure}
	\includegraphics[width=0.5\textwidth]{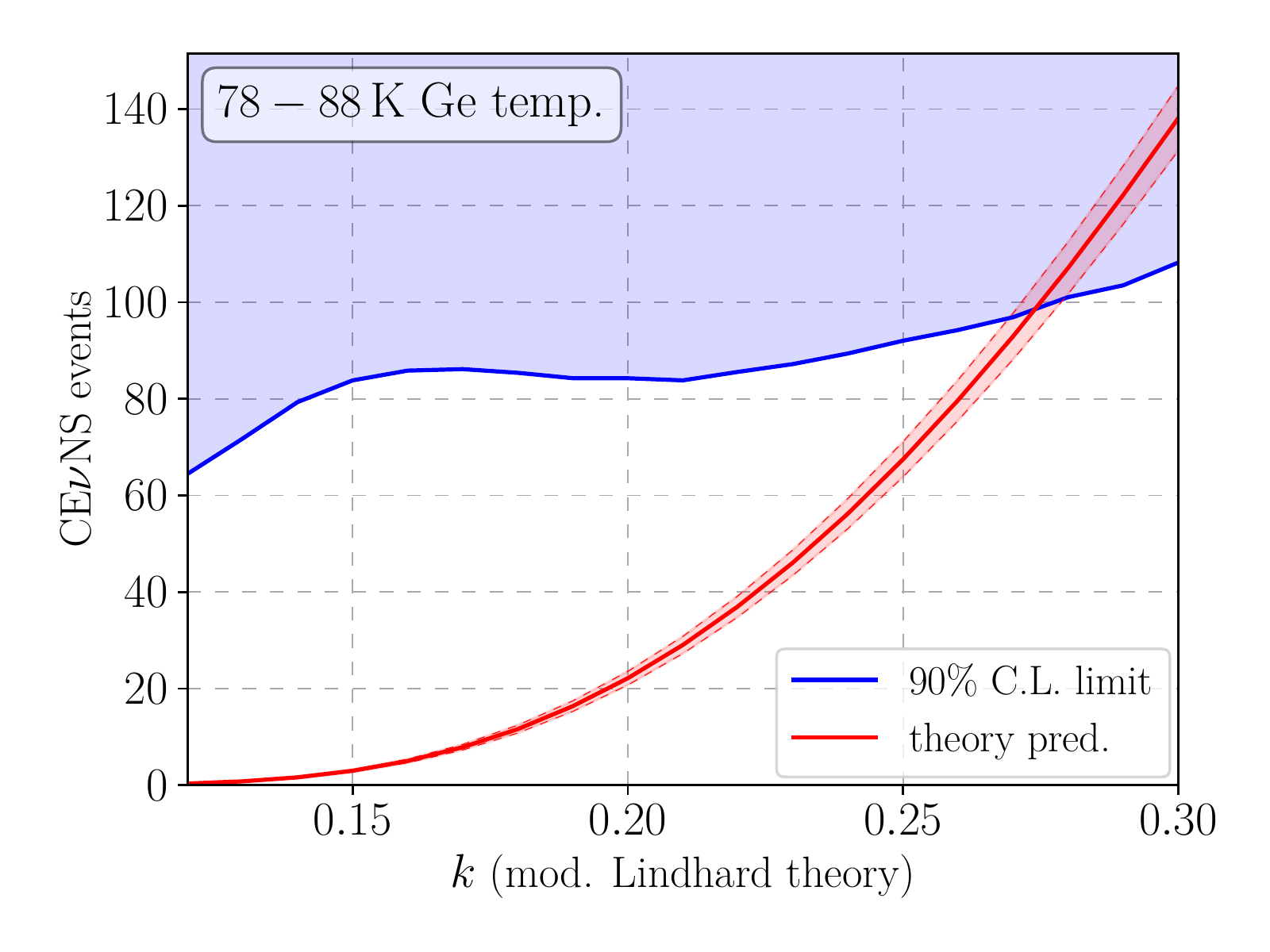}
	\caption{The upper limit (90\% C.L.) on the number of CE$\nu$NS counts (blue curve) is shown as a function of the quenching parameter. For comparison the predicted count rate is plotted in red.}
	\label{fig:4}
\end{figure}

With the current regions of interest no hint for a CE$\nu$NS signal can be observed in the data yet. Therefore, we extracted an upper limit (U.L.) on the observed number of CE$\nu$NS events applying a likelihood ratio test. In this test, the likelihood is scanned over a range of $s$ parameters and the hypothesis of a signal is always compared to the null hypothesis of no signal. By using the best fit of the other parameters, this scan can be converted to a scan over the number of signal counts. Employing  toy MC simulations, we can determine the distribution of the test statistic and extract a limit on the number of signal events as shown in Figure~\ref{fig:4}. The test statistic follows approximately a $\chi^2$ distribution as expected from Wilks' theorem. The resulting 90\,\%~C.L.~limit on the number of signal events for a $k$ parameter of 0.18 (close to the value of $k$ determined in~\cite{Scholz:2016}) is 85 in the ROIs. This corresponds to less than 0.34\,events kg$^{-1}\cdot$d$^{-1}$ during reactor ON periods. From the predictions of the SM and our reactor model we would expect in total $11.6\pm0.8$ events in the ROIs at the same quenching factor, about a factor 7 below the experimental 90\,\%~C.L.~limit. In Figure~\ref{fig:4}, the upper limits on the number of signal events are shown as a function of the quenching parameter k. Comparing the upper limits with the expectation, $k$ values larger than 0.27 are disfavored by the \textsc{Conus} data.

The reactor in Brokdorf will finish operation by the end of 2021. With additional run periods, in particular including the substantial additional reactor OFF time after 2021 the statistical uncertainty will shrink significantly. Moreover, further measures toward a lower energy threshold and an improved signal-to-background ratio should both help to boost our sensitivity. The ability of \textsc{Conus} to detect a CE$\nu$NS signal after these upgrades and with additional data depends strongly on the true value of the Ge quenching parameter. This can be seen from the strong dependence of the signal expectation on the $k$ parameter as illustrated in Figure~\ref{fig:4}. For a reliable sensitivity projection and an accurate signal prediction additional precision measurements of the quenching in Ge crystals at the relevant recoil energies and temperature are mandatory.  

We thank all the technical and administrative staff who helped building the experiment, in particular the MPIK workshops and Mirion Technologies (Canberra) in Lingolsheim. We express our gratitude to the Preussen Elektra GmbH for great support and for hosting the \textsc{Conus} experiment. We thank Dr.~S.~Schoppmann (MPIK) for assistance on the analysis and Dr.~M.~Seidl (Preussen Elektra) for providing simulation data on the fission rate evolution over a reactor cycle. The \textsc{Conus} experiment is supported financially by the Max Planck Society (MPG), T.~Rink by the German Research Foundation (DFG) through the research training group GRK 1940, and together with J.~Hakenm\"uller by the IMPRS-PTFS.

\bibliographystyle{ieeetr}
\bibliography{references}

\end{document}